\documentclass[11pt]{article}
\begin{document}
\date{}
\begin{center}
{\Large {\bf FEDOSOV SUPERMANIFOLDS: BASIC PROPERTIES}}
\end{center}

\begin{center}
{\Large {\bf AND THE DIFFERENCE IN EVEN AND ODD CASES }}
\end{center}

{\Large
\begin{center}
\medskip
{\sc B.~Geyer}$^{\ a)}$\footnote{E-mail: geyer@itp.uni-leipzig.de},
and
{\sc P.M.~Lavrov}$^{\ a), b)}$\footnote{E-mail: lavrov@tspu.edu.ru;
lavrov@itp.uni-leipzig.de}\\

\vspace{1cm}
{\normalsize\it $^{a)}$ Center of Theoretical Studies, Leipzig University,\\
Augustusplatz 10/11, D-04109 Leipzig, Germany}

\vspace{.4cm}

{\normalsize\it $^{b)}$ Tomsk State Pedagogical University,
634041 Tomsk, Russia}
\end{center}
}
\vspace{.5cm}

\begin{quotation}
\setlength{\baselineskip}{12pt} \normalsize \noindent
 We study
basic properties of supermanifolds endowed with an even (odd)
symplectic structure and a connection respecting this symplectic
structure. Such supermanifolds can be considered as generalization
of Fedosov manifolds to the supersymmetric case. Choosing an
appropriate definition of inverse (second-rank) tensor fields on
supermanifolds we consider the symmetry behavior of tensor fields
as well as the properties of the symplectic curvature and of the
Ricci tensor on even (odd) Fedosov supermanifolds. We show that
for odd Fedosov supermanifolds the scalar curvature, in general,
is non-trivial while for even Fedosov supermanifolds it necessarly
vanishes.
\end{quotation}

\bigskip
\begin{center}
{\bf 1. INTRODUCTION}
\end{center}

The methods of modern differential geometry have become an
universal tool of theoretical physics starting from the
recognition of their crucial role in general relativity. In
general relativity the basic objects are Riemannian manifolds
equipped with a symmetric connection respecting a metric tensor
field. The formulation of classical mechanics involves symplectic
manifolds, i.e., manifolds endowed with a non-degenerate closed
2-form used for the construction of the Poisson bracket. It has
also been realized that the so-called deformation quantization
\cite{F} can be formulated in terms of symplectic manifolds
equipped with a symmetric connection respecting the symplectic
structure (Fedosov manifolds) \cite{fm}.

The discovery of supersymmetric particle theories as well as of
supergravity has introduced into modern quantum field theory a
number of applications of differential geometry which are based on
the notion of supermanifolds, first proposed and analyzed by
Berezin \cite{Ber} (see also \cite{DeWitt}). In these cases, one
has to equip the supermanifolds with a suitable connection. The
general consideration of the well-known Batalin-Vilkovisky
quantization method \cite{bv} involves an odd symplectic
supermanifold, i.e., a supermanifold endowed with an odd
symplectic structure \cite{geom}. In some specific considerations
of modern gauge field theory (see, e.g., \cite{bt,gl}) one
introduced also even symplectic supermanifolds equipped with a
flat connection respecting a given symplectic structure.

The goal of the present paper is to study the symmetry properties
of tensor fields on supermanifolds, as well as the properties of
the curvature tensor for arbitrary even (odd) symplectic
supermanifolds endowed with a symmetric connection respecting a
given symplectic structure.

The paper is organised as follows. In Sect.~2, we  remind  the
definition of tensor fields on  supermanifolds. In Sect.~3, we
consider affine connections on a supermanifold and their curvature
tensors. In Sect.~4, we discuss relations between even (odd)
symplectic supermanifolds and even (odd) Poisson supermanifolds.
In Sect.~5, we present the notion of even (odd) Fedosov
supermanifolds and of even (odd) symplectic curvature tensors. In
Sect.~6, we study the Ricci tensor constructed from the symplectic
curvature and the scalar curvature which is non-trivial for odd
Fedosov supermanifolds. In Sect.~7, we give a short summary.

We use the condensed notation suggested by DeWitt. Derivatives
with respect to the coordinates $x^i$ are understood as acting
from the left and for them the notation $\partial_i A={\partial
A}/{\partial x^i}$ is used. Right derivatives with respect to
$x^i$ are labelled by the subscript $"r"$ or the notation
$A_{,i}={\partial_r A}/{\partial x^i}$ is used. The Grassmann
parity of any quantity $A$ is denoted by $\epsilon (A)$.
\\

\date{}
\begin{center}
{\bf 2. TENSOR  FIELDS ON SUPERMANIFOLDS}
\end{center}

In this Chapter we review explicitly some basic definitions and
simple relations of tensor analysis on supermanifolds which are
useful in order to avoid elementary pitfalls in the course of the
computations. Thereby, we adopt the conventions of
DeWitt~\cite{DeWitt}.

Let the variables $x^i, \epsilon(x^i)=\epsilon_i$ be local
coordinates of a supermanifold $M, dim M=N,$ in the vicinity of a
point $P$. Let the sets $\{e_i\}$ and  $\{e^i\}$ be coordinate
bases in the tangent space $T_PM$ and the cotangent space
$T^*_PM$, respectively. If one goes over to another set ${\bar
x}^{i}={\bar x}^i(x)$ of local coordinates the basis vectors in
$T_PM$ and $T^*_PM$ transform as follows:
\begin{eqnarray}
\label{vec}
 {\bar e}_i=e_j \frac{\partial_r x^j}{\partial {\bar x}^i},
 \quad
{\bar e}^i=e^j \frac{\partial {\bar x}^i}{\partial x^j}.
\end{eqnarray}
For the transformation matrices the following relations hold:
\begin{eqnarray}
\label{unitJ}
 \frac{\partial_r {\bar x}^i}{\partial x^k}
 \frac{\partial_r x^k}{\partial {\bar x}^j}=\delta^i_j,
 \quad
 \frac{\partial x^k}{\partial {\bar x}^j}
 \frac{\partial {\bar x}^i}{\partial x^k}=\delta^i_j,
 \quad
 \frac{\partial_r x^i}{\partial {\bar x}^k}
 \frac{\partial_r {\bar x}^k}{\partial  x^j}=\delta^i_j,
 \quad
 \frac{\partial {\bar x}^k}{\partial  x^j}
 \frac{\partial  x^i}{\partial {\bar x}^k}=\delta^i_j.
\end{eqnarray}

A tensor field of type $(n,m)$ with rank $n+m$ is defined as a
geometric object which, in each local coordinate system
$(x)=(x^1,...,x^N)$, is given by a set of functions with $n$ upper
and $m$ lower indices obeying definite transformation rules. For
example, let that set of functions be given either as
$T^{i_1...i_n}_{\;\;\;\;\;\;\;\;\;\;j_1...j_m}(x)$ or as
$T^{\;\;\;\;\;\;\;\;\;\;i_1...i_n}_{j_1...j_m}(x)$ with Grassmann
parity $\epsilon(T^{\;\;\;\;\;\;\;\;\;\;i_1...i_n}_{j_1...j_m})=
\epsilon(T^{i_1...i_n}_{\;\;\;\;\;\;\;\;\;\;j_1...j_m})=
\epsilon(T)+\epsilon_{i_1}+\cdot\cdot\cdot + \epsilon_{i_n}+
\epsilon_{j_1}+\cdot\cdot\cdot +\epsilon_{j_m}$. Then the
transformation rules under a change of coordinates,
$(x)\rightarrow ({\bar x})$, are given as follows (assuming the
usual convention on the ordering of transformation matrices):
\begin{eqnarray}
\label{tenzor} {\bar
T}^{i_1...i_n}_{\;\;\;\;\;\;\;\;\;\;\;j_1...j_m}&=&
T^{l_1...l_n}_{\;\;\;\;\;\;\;\;\;\;k_1...k_m} \frac{\partial_r
x^{k_m}}{\partial {\bar x}^{j_m}}\cdot\cdot\cdot \frac{\partial_r
x^{k_1}}{\partial {\bar x}^{j_1}} \frac{\partial {\bar
x}^{i_n}}{\partial x^{l_n}}\cdot\cdot\cdot
\frac{\partial {\bar x}^{i_1}}{\partial x^{l_1}}\\
\nonumber && \times
(-1)^{\left(\sum\limits^{m-1}_{s=1}\sum\limits^{m}_{p=s+1}
\epsilon_{j_p}(\epsilon_{j_s}+\epsilon_{k_s}) +
\sum\limits^{n}_{s=1}\sum\limits^{m}_{p=1}\epsilon_{j_p}
(\epsilon_{i_s}+\epsilon_{l_s})+
\sum\limits^{n-1}_{s=1}\sum\limits^{n}_{p=s+1}
\epsilon_{i_p}(\epsilon_{i_s}+\epsilon_{l_s})\right)}
\end{eqnarray}
or
\begin{eqnarray}
\label{tenzor1} {\bar
T}^{\;\;\;\;\;\;\;\;\;\;i_1...i_n}_{j_1...j_m}
&=&
T^{\;\;\;\;\;\;\;\;\;\;l_1...l_n}_{k_1...k_m}
\frac{\partial {\bar
x}^{i_n}}{\partial x^{l_n}}\cdot\cdot\cdot
\frac{\partial {\bar x}^{i_1}}{\partial x^{l_1}}
\frac{\partial_r
x^{k_m}}{\partial {\bar x}^{j_m}}\cdot\cdot\cdot \frac{\partial_r
x^{k_1}}{\partial {\bar x}^{j_1}} \\
\nonumber && \times
(-1)^{\left(\sum\limits^{n-1}_{s=1}\sum\limits^{n}_{p=s+1}
\epsilon_{i_p}(\epsilon_{i_s}+\epsilon_{l_s}) +
\sum\limits^{n}_{s=1}\sum\limits^{m}_{p=1}\epsilon_{i_s}
(\epsilon_{k_p}+\epsilon_{j_p})+
\sum\limits^{m-1}_{s=1}\sum\limits^{m}_{p=s+1}
\epsilon_{j_p}(\epsilon_{k_s}+\epsilon_{j_s})\right)}\,.
\end{eqnarray}

For vector fields, $T^i$, and covector fields, $T_i$, the
transformation rule is obvious:
\begin{eqnarray}
\label{formvec}
 {\bar T}^i= T^n\frac{\partial {\bar x}^i}{\partial x^n}\,, 
 \qquad
 {\bar T}_i= T_n\frac{\partial_r x^n}{\partial{\bar x}^i}\,.
\end{eqnarray}
For second-rank tensor fields of different type, (\ref{tenzor})
and (\ref{tenzor1}) imply the transformation rules
\begin{eqnarray}
\label{formup}
{\bar T}^{ij}&=&
T^{mn}\frac{\partial {\bar x}^j}{\partial x^n}
\frac{\partial {\bar x}^i}{\partial x^m}
(-1)^{\epsilon_j(\epsilon_i+\epsilon_m)},\\
\label{form}
{\bar T}_{ij}&=&
T_{mn}\frac{\partial_r x^n}{\partial {\bar x}^j}
\frac{\partial_r x^m}{\partial {\bar x}^i}
(-1)^{\epsilon_j(\epsilon_i+\epsilon_m)},\\
\label{form1} {\bar T}^i_{\;\;j}&=& T^m_{\;\;\;n}\frac{\partial_r
x^n}{\partial {\bar x}^j} \frac{\partial {\bar x}^i}{\partial x^m}
(-1)^{\epsilon_j(\epsilon_i+\epsilon_m)},\\
\label{form2} {\bar T}_i^{\;\;j}&=& T_m^{\;\;\;n}\frac{\partial
{\bar x}^j}{\partial x^n} \frac{\partial_r x^m}{\partial {\bar
x}^i} (-1)^{\epsilon_j(\epsilon_i+\epsilon_m)}\,.
\end{eqnarray}
Note that the unit matrix $\delta^i_j$ is connected with unit
tensor fields $\delta^i_{\;j}$ and $\delta^{\;\;i}_j$,
transforming according to (\ref{form1}) and (\ref{form2}), as
follows
\begin{eqnarray}
\label{unit}
\delta^i_j=\delta^i_{\;j}=(-1)^{\epsilon_i}\;\delta^{\;\;i}_j =
(-1)^{\epsilon_j}\;\delta^{\;\;i}_j.
\end{eqnarray}

From a tensor field of type $(n,m)$ with rank $n+m$, where $n\neq
0, \;m\neq 0$, one can construct a tensor field of type $(n-1,m-1)$
with rank $n+m-2$ by the contraction of an upper and a lower index
by the rules
\begin{eqnarray}
\label{tencontr} &&
 T^{i_1...i_{s-1}\;i \;i_{s+1}...i_n}
 _{\;\;\;\;\;\;\;\;\;\;\;\;\;\;\;\;\;\;\;\;\;\;\;\;\;\;\;\;\;\;\;
 j_1...j_{q-1}\;i\;j_{q+1}...j_m}\,
 (-1)^{\epsilon_i(\epsilon_{i_{s+1}}+\cdot\cdot\cdot +
 \epsilon_{i_n} + \epsilon_{j_1}+\cdot\cdot\cdot
 +\epsilon_{j_{q-1}}+1)},\\
\nonumber
\\
 \label{tencontr1} &&
 T_{ j_1...j_{q-1}\;i\;j_{q+1}...j_m}
 ^{\;\;\;\;\;\;\;\;\;\;\;\;\;\;\;\;\;\;\;\;\;\;\;\;\;\;\;\;\;\;\;
 i_1...i_{s-1}\;i \;i_{s+1}...i_n}\;
 (-1)^{\epsilon_i(\epsilon_{i_{s+1}}+\cdot\cdot\cdot +
 \epsilon_{i_n} + \epsilon_{j_1}+\cdot\cdot\cdot
 +\epsilon_{j_{q-1}})}.
\end{eqnarray}
In particular, for the tensor fields of type $(1,1)$ the
contraction leads to the supertraces,
\begin{eqnarray}
\label{sc1} T^i_{\;\;i}\;(-1)^{\epsilon_i}
 \quad {\rm and}\quad T^{\;\;i}_i.
\end{eqnarray}

Having in mind the index shifting rules of DeWitt~\cite{DeWitt},
\begin{eqnarray}
\label{shift}
 {T}^i \;=\; ^iT \;(-1)^{\epsilon(T)\epsilon_i}
 \qquad {\rm and} \qquad
 {T}_i \;=\; _iT\; (-1)^{\epsilon(T)\epsilon_i+\epsilon_i},
\end{eqnarray}
and similarly for tensors of higher rank, let us construct from
two tensor fields $U^{i_1...i_n k}$ and $V_{l j_1...j_m}$ new
tensor fields by
\begin{eqnarray}
\label{contr1}
 (-1)^{\epsilon(V)(\epsilon_{i_{1}}+\cdots +
 \epsilon_{i_n}+\epsilon_{k}) + \epsilon_{k}}\;
 U^{i_1...i_n k}\;V_{k j_1...j_m}&\equiv&
 T^{i_1...i_n}_{\;\;\;\;\;\;\;\;\;\;\;j_1...j_m}\,,\\
\label{contr2}
 (-1)^{\epsilon(U)(\epsilon_{j_{1}}+\cdots +
 \epsilon_{j_m}+\epsilon_{k})}\;
 V_{j_1...j_m k}\;U^{k i_1...i_n }
 &\equiv& 
 \widehat T^{\;\;\;\;\;\;\;\;\;\;\;i_1...i_n}_{j_1...j_m}\;,
\end{eqnarray}
which really transform according to (\ref{tenzor}) and
(\ref{tenzor1}), respectively. For example, from vector and
covector fields, $U^i$ and $V_i$, one gets a scalar field, i.e., a
tensor field of rank zero,
\begin{eqnarray}
\label{sc}
 (-1)^{\epsilon_i(\epsilon(V)+1)}\;U^i\;V_i =
 (-1)^{\epsilon(U)\epsilon(V) + \epsilon_i\epsilon(U)}\;V_i\;U^i\,,
\end{eqnarray}
being an invariant, and from two second rank tensor fields,
$U^{ij}$ and $V_{ij}$, we obtain
\begin{eqnarray}
\label{contr}
(-1)^{(\epsilon_i+\epsilon_k)\epsilon(V)+\epsilon_k}\;U^{ik}\;V_{kj}
\quad {\rm and} \quad
(-1)^{(\epsilon_i+\epsilon_k)\epsilon(U)}\;V_{ik}\;U^{kj},
\end{eqnarray}
transforming according to (\ref{form1}) and (\ref{form2}),
respectively, but after a further contraction, on gets the scalar
\begin{eqnarray}
\label{contr3}
(-1)^{(\epsilon_i+\epsilon_k)(\epsilon(V)+1)}\;U^{ik}\;V_{ki} =
(-1)^{\epsilon(U)\epsilon(V)+
(\epsilon_i+\epsilon_k)\epsilon(U)}\;V_{ik}\;U^{ki}\,.
\end{eqnarray}
Furthermore, taking into account (\ref{contr1}) and
(\ref{contr2}), the unique inverse of a (non-degenerate) second
rank tensor field of type (2,0) will be defined as follows:
\footnote{Taking into account the index shifting rules
(\ref{shift}) and the
properties (\ref{unit}) the inverse tensor obeys $ %
^iT^{k}\;_k(T^{-1})_{j}\; = \;\delta^i_{j} \;=\;
_j(T^{-1})_{k}\;^kT^{i} \,. $ 
In principle, also a factor $(-1)^{\epsilon(T)}$, related to the
exchange of $T$ and $T^{-1}$, could be included into the relation
(\ref{invers2}); but then, in general, left and right inverse
become different.}
\begin{eqnarray}
\label{invers1}
(-1)^{(\epsilon_i+\epsilon_k)\epsilon(T)+\epsilon_k}\;T^{ik}\;(T^{-1})_{kj}
&=& \delta^i_{\;j}\,,\\
\label{invers2}
(-1)^{(\epsilon_j+\epsilon_k)\epsilon(T)}\;(T^{-1})_{jk}\;T^{ki}
&=& \delta_j^{\;\;i}\,,
\\
\nonumber \quad
 \epsilon(T^{-1}_{ij})=\epsilon(T^{ij})=
 \epsilon(T)+\epsilon_i+\epsilon_j\,,&&
\end{eqnarray}
and correspondingly for tensor fields of type (0,2).


Let us remark that the inclusion of the correct sign factors into
the definitions of contractions, (\ref{tencontr}) and
(\ref{tencontr1}), and of the inverse tensors, (\ref{invers1}) and
(\ref{invers2}), is essential. Namely,
let us consider a second rank tensor field of type (2,0) obeying
the property of generalized (anti)symmetry,
\begin{eqnarray}
\label{sym}
 T_\pm^{ij}&=&\pm(-1)^{\epsilon_i\epsilon_j}T_\pm^{ji}. 
\end{eqnarray}
Obviously, that property is in agreement with the transformation
law (\ref{formup}),
\begin{eqnarray}
\nonumber
&&{\bar T}_\pm^{ij}= T_\pm^{mn}\frac{\partial {\bar
x}^j}{\partial x^n} \frac{\partial {\bar x}^i}{\partial x^m}
(-1)^{\epsilon_j(\epsilon_i+\epsilon_m)}= \pm T_\pm^{nm}
\frac{\partial {\bar x}^i}{\partial x^m} \frac{\partial {\bar
x}^j}{\partial x^n} (-1)^{\epsilon_i\epsilon_n} =\pm
(-1)^{\epsilon_i\epsilon_j}{\bar T}_\pm^{ji}.
\end{eqnarray}
Thus, the notion of generalized (anti)symmetry of a tensor field
of type (2,0) is invariantly defined in any coordinate system.

Now, suppose that $T_\pm^{ij}$ is non-degenerate, thus allowing
for the introduction of the corresponding inverse tensor fields of
type (0,2) according to (\ref{invers1}) and (\ref{invers2}). From
(\ref{sym}) one gets
\begin{eqnarray}
\label{inversym}
(T^{-1}_\pm)_{ij}=\pm(-1)^{\epsilon_i\epsilon_j+\epsilon(T)}(T^{-1}_\pm)_{ji}\,,
\end{eqnarray}
and, as it should be, also this generalized (anti)symmetry is
invariantly defined.

However, if the inverse tensor field had been defined naively
according to
\begin{eqnarray}
\label{inver1} T^{ik}_\pm\;(\widetilde T^{-1}_\pm)_{kj} =
\delta^i_{j}\,, \qquad
(\widetilde T^{-1}_\pm)_{ik}\;T^{kj}_\pm =\delta_i^{j}\,,
\end{eqnarray}
then, instead of Eq.~(\ref{inversym}) one obtains
\begin{eqnarray}
\label{wrong}
 (\widetilde T^{-1}_\pm)_{ij}=\mp(-1)^{(\epsilon_i+1)(\epsilon_j+1)
 +\epsilon(T)}(\widetilde T^{-1}_\pm)_{ji},\quad
 \epsilon((\widetilde T^{-1}_\pm)_{ij})= 
 \epsilon(T_\pm)+\epsilon_i+\epsilon_j\,.
\end{eqnarray}
However, that symmetry property of $(\widetilde T^{-1}_\pm)_{ij}$
has no invariant meaning. Namely, the transformed field $
({\overline{\widetilde T^{-1}_\pm}})_{ij}$, being determined
according to the rule (\ref{form}),
does not have the symmetry (\ref{wrong}). The reason is, that
(\ref{inver1}) does not introduce $(\widetilde T^{-1}_\pm)_{ij}$
as a tensor field on a supermanifold.

This observations have to be taken into account later on when
considering Poisson and symplectic supermanifolds.
Furthermore, among supermatrices with all possible generalized
symmetry properties (see, e.g.,~\cite{GT}) only supermatrices with
the properties (\ref{sym}) have an invariant meaning. Therefore,
differential geometry on supermanifolds should be developed solely
on the basis of tensor fields that possess the properties of the
generalized (anti)symmetry.
\\

\begin{center}
{\bf 3. AFFINE CONNECTIONS ON SUPERMANIFOLDS AND CURVATURE}
\end{center}

As in the case of tensor analysis on manifolds, on a supermanifold
$M$ one can introduce the covariant derivation (or affine
connection) as a mapping $\nabla$ (with components $\nabla_i,\,
\epsilon(\nabla_i)= \epsilon_i$) from the set of tensor fields on
$M$ to itself by the requirement that it should be a tensor
operation acting from the right and adding one more lower index
and, when it is possible locally to introduce Cartesian
coordinates on $M$, that it should reduce to the usual
(right--)differentiation.

Let us first discuss the latter case, with $(x)$ being a Cartesian
coordinate system and $({\bar x})$ an arbitrary one. Let us
consider a vector field $T^i$. Then, in the system $(x)$, we have
\begin{eqnarray}
\nonumber T^i\nabla_j=T^i_{\;\;,j}\,,
\end{eqnarray}
and in the coordinate system $({\bar x})$, by virtue of
(\ref{form1}) and (\ref{formvec}), we obtain
\begin{eqnarray}
\nonumber {\bar T}^i{\bar \nabla}_j= (T^m
_{\;\;\;,n})\frac{\partial_r x^n}{\partial {\bar x}^j}
\frac{\partial {\bar x}^i}{\partial x^m}
(-1)^{\epsilon_j(\epsilon_i+\epsilon_m)} =
{\bar T}^i_{\;\;,j} + {\bar T}^k\Gamma^i_{\;\;kj}
(-1)^{\epsilon_k(\epsilon_i+1)}\,;
\end{eqnarray}
here, $\Gamma^i_{\;\;jk}$ are the affine connection components (or
Christoffel symbols),
\begin{eqnarray}
\label{Cris} \Gamma^i_{\;\;kj}= \frac{\partial_r {\bar
x}^i}{\partial x^m} \frac{\partial^2_r x^m}{\partial {\bar
x}^k\partial {\bar x^j}}\,.
\end{eqnarray}
By definition they possess the property of generalized symmetry
w.r.t. the lower indices,
\begin{eqnarray}
\label{Crisp}
\Gamma^i_{\;\;jk}= (-1)^{\epsilon_j\epsilon_k}\Gamma^i_{\;\;kj}.
\end{eqnarray}

Similarly, the action of the covariant derivative on covector
fields $T_i$ is given by
\begin{eqnarray}
\nonumber {\bar T}_i{\bar \nabla}_j= {\bar T}_{i,j} +{\bar
T}_k{\widetilde\Gamma}^k_{\;\;ij} \qquad {\rm with} \qquad
{\widetilde\Gamma}^k_{\;\;ij}=
(-1)^{\epsilon_j(\epsilon_i+\epsilon_l)} \frac{\partial^2_r {\bar
x}^k}{\partial x^l\partial x^m} \frac{\partial_r x^m}{\partial
{\bar x}^j} \frac{\partial_r x^l}{\partial {\bar x}^i}\,.
\end{eqnarray}
Using the first of relations (\ref{unitJ}), one establishes
straightforwardly that
\begin{eqnarray}
\nonumber {\widetilde\Gamma}^k_{\;\;ij}=-\Gamma^k_{\;\;ij}\,,
\end{eqnarray}
and therefore
\begin{eqnarray}
\nonumber {\bar T}_i{\bar \nabla}_j= {\bar T}_{i,j} -{\bar
T}_k\Gamma^k_{\;\;ij}\,.
\end{eqnarray}

In general, on arbitrary supermanifolds $M$, the Christoffel
symbols are not necessarily given by partial derivatives with
respect to the coordinates. However, also
$M$.
for arbitrary supermanifolds the covariant derivative $\nabla $
(or connection $\Gamma$) is defined through the (right--)
differentiation and the separate contraction of upper and lower
indices with the connection components analogous to the case of
vector and co-vector fields. More explicitly, they are given as
local operations acting on scalar, vector and co-vector fields
by the rules
\begin{eqnarray}
\label{scal} T\,\nabla_i&=&T_{,i}\,,
\\
\label{vector} T^i\,\nabla_j&=&T^i_{\;\;,j}+ T^k\Gamma^i_{\;\;kj}
(-1)^{\epsilon_k(\epsilon_i+1)}\,,
\\
T_i\,\nabla_j&=&T_{i,j}-T_k\Gamma^k_{\;\;ij}\,,
\end{eqnarray}
and on second-rank tensor fields of type $(2,0), (0,2)$ and
$(1,1)$ by the rules
\begin{eqnarray}
{T}^{ij}\,{\nabla}_k&=& {T}^{ij}_{\;\;\;,k} +
{T}^{il}\,\Gamma^j_{\;\;lk}(-1)^{\epsilon_l(\epsilon_j+1)}+
{T}^{lj}\,\Gamma^i_{\;\;lk}
(-1)^{\epsilon_i\epsilon_j+\epsilon_l(\epsilon_i+\epsilon_j+1)}\,,\\
{T}_{ij}\,{\nabla}_k&=& {T}_{ij,k} -
{T}_{il}\,\Gamma^l_{\;\;jk}-
{T}_{lj}\,\Gamma^l_{\;\;ik}
(-1)^{\epsilon_i\epsilon_j+\epsilon_l\epsilon_j}\,,\\
{T}^i_{\;\;j}\,{\nabla}_k&=& { T}^i_{\;\;j,k} -
{T}^i_{\;\;l}\,\Gamma^l_{\;\;jk} +
{T}^l_{\;\;j}\,\Gamma^i_{\;\;lk}
(-1)^{\epsilon_i\epsilon_j+\epsilon_l(\epsilon_i+\epsilon_j+1)}\,.
\end{eqnarray}
Similarly, the action of the covariant derivative on a tensor
field of any rank and type is given in terms of their tensor
components, their ordinary derivatives and the connection
components.

The affine connection components do not transform as mixed tensor
fields, instead they obtain an additional inhomogeneous term:
\begin{eqnarray}
{\bar\Gamma}^i_{\;\;jk}= (-1)^{\epsilon_n(\epsilon_m+\epsilon_j)}
 \frac{\partial_r \bar x^i}{\partial x^l}\Gamma^l_{\;\;mn}
 \frac{\partial_r  x^m}{\partial \bar x^j}
 \frac{\partial_r  x^n}{\partial \bar x^k}
 +
 \frac{\partial_r \bar x^i}{\partial x^m}
 \frac{\partial_r^2  x^m}{\partial \bar x^j \partial \bar x^k}\,.
\end{eqnarray}
In general, the connection components $\Gamma^i_{\;\;jk}$ do not
have the property of (generalized) symmetry w.r.t. the lower
indices. The deviation from this symmetry is the torsion,
\begin{eqnarray}
T^i_{\;\;jk} := \Gamma^i_{\;\;jk} -
(-1)^{\epsilon_j\epsilon_k}\Gamma^i_{\;\;kj}\,,
\end{eqnarray}
which transforms as a tensor field.
 If the supermanifold $M$ is torsionless, i.e., if the Christoffel symbols
obey the relation (\ref{Crisp}), then one says that a symmetric
connection is defined on $M$. Here, with the aim of studying
Fedosov supermanifolds, we consider only symmetric connections.

The Riemannian tensor field $R^i_{\;\;mjk}$, according to
Ref.~\cite{DeWitt}, is defined in a coordinate basis by the action
of the commutator of covariant derivatives, $[\nabla_i,\nabla_j]=
\nabla_i\nabla_j-(-1)^{\epsilon_i\epsilon_j}\nabla_j\nabla_i$, on
a vector field $T^i$ as follows:
\begin{eqnarray}
T^i[\nabla_j,\nabla_k]=-(-1)^{\epsilon_m(\epsilon_i+1)}
T^mR^i_{\;\;mjk}.
\end{eqnarray}
A straightforward calculation yields 
\begin{eqnarray}
\label{R} R^i_{\;\;mjk}=-\Gamma^i_{\;\;mj,k}+
\Gamma^i_{\;\;mk,j}(-1)^{\epsilon_j\epsilon_k}+
\Gamma^i_{\;\;jn}\Gamma^n_{\;\;mk}(-1)^{\epsilon_j\epsilon_m}-
\Gamma^i_{\;\;kn}\Gamma^n_{\;\;mj}
(-1)^{\epsilon_k(\epsilon_m+\epsilon_j)}.
\end{eqnarray}
The Riemannian tensor field possesses the following generalized
antisymmetry property,
\begin{eqnarray}
\label{Rsym}
R^i_{\;\;mjk}=-(-1)^{\epsilon_j\epsilon_k}R^i_{\;\;mkj}\,;
\end{eqnarray}
furthermore, it obeys the (super) Jacobi identity,
\begin{eqnarray}
\label{Rjac} (-1)^{\epsilon_m\epsilon_k}R^i_{\;\;mjk}
+(-1)^{\epsilon_j\epsilon_m}R^i_{\;\;jkm}
+(-1)^{\epsilon_k\epsilon_j}R^i_{\;\;kmj}\equiv 0\,.
\end{eqnarray}
Using the (super) Jacobi identity for the covariant derivatives,
\begin{eqnarray}
\label{}
[\nabla_i,[\nabla_j,\nabla_k]](-1)^{\epsilon_i\epsilon_k}+
[\nabla_k,[\nabla_i,\nabla_j]](-1)^{\epsilon_k\epsilon_j}+
[\nabla_j,[\nabla_k,\nabla_i]](-1)^{\epsilon_i\epsilon_j}\equiv
0\,,
\end{eqnarray}
one obtains the (super) Bianchi identity,
\begin{eqnarray}
\label{BI} (-1)^{\epsilon_i\epsilon_j}R^n_{\;\;mjk;i}
+(-1)^{\epsilon_i\epsilon_k}R^n_{\;\;mij;k}
+(-1)^{\epsilon_k\epsilon_j}R^n_{\;\;mki;j}\equiv 0\,,
\end{eqnarray}
with the notation $R^n_{\;\;mjk;i}:\,=R^n_{\;\;mjk}\nabla_i$.
\\

\begin{center}
{\bf 4. SYMPLECTIC AND POISSON SUPERMANIFOLDS}
\end{center}

Suppose now that we are given a supermanifold $M$ of an even
dimension, ${\rm dim}\, M=2n$. Let $\omega$ be an even,
$\epsilon(\omega)=0$, (resp. odd, $\epsilon(\omega)=1$)
non-degenerate exterior 2-form on $M$. Then, the pair $(M,\omega)$
is called an even (resp. odd) almost symplectic supermanifold; it
is called an even (resp. odd) symplectic supermanifold if $\omega$
is closed, $d \omega = 0$.

In an arbitrary coordinate basis on $M$, $\omega$ can be written
as \footnote{Let us remark that in the literature various other
definitions occur which seemingly differ from our one. However,
the difference may be resolved by partially including (overall)
sign factors into $\omega$ and changing the transformation rules,
e.g., re-interpreting $\omega_{ij}$ as $\;_j\omega_i\;$.}
\begin{eqnarray}
\label{A13} \omega &=& \omega_{ij}\;dx^j\wedge dx^i 
\\ 
\omega_{ij}&=&-(-1)^{\epsilon_i\epsilon_j}\;\omega_{ji}, \;\quad
dx^i\wedge dx^j \;=\; -(-1)^{\epsilon_i\epsilon_j}\;dx^j\wedge dx^i,\\
\nonumber
\epsilon(\omega_{ij})&=&\epsilon(\omega)+\epsilon_i+\epsilon_j,\quad
\epsilon(dx^i\wedge dx^j)\;=\;\epsilon_i+\epsilon_j,
\end{eqnarray}
where $\omega_{ij}$ is a second-rank tensor field of type $(0,2)$,
and the wedge product $dx^i\wedge dx^j$ is a second-rank tensor
field of type $(2,0)$. Note, that the symmetry properties of
$\omega_{ij}$ and of the wedge product $dx^j\wedge dx^i$ are
respected by the transformation laws (\ref{form}) and
(\ref{formup}), respectively, and, taking into account
(\ref{contr1}), one concludes that $\omega$ is invariant under a
change of local coordinates, ${\bar\omega}=\omega$.

The exterior derivative, defined for a 2-form according to
\begin{eqnarray}
\label{A14} d\omega = \omega_{ij,k}\;dx^k\wedge dx^j\wedge dx^i,
\quad d^2\omega=0\,,
\end{eqnarray}
is also invariant under a change of local coordinates,
$\overline{d\omega} = d\omega$. The requirement of closure,
$d\omega=0$, leads to the following identity for $\omega_{ij}$:
\begin{eqnarray}
\label{A12} \omega_{ij,k}(-1)^{\epsilon_i\epsilon_k}+cycle
(i,j,k)\equiv 0 \quad\longleftrightarrow\quad
\partial_i\omega_{jk}(-1)^{\epsilon_i(\epsilon(\omega)+1+\epsilon_k)}
+ cycle (i,j,k)\equiv 0.
\end{eqnarray}

Suppose now that the tensor field $\omega_{ij}$ is non-degenerate.
Taking into account the relations (\ref{unit}) and the definitions
of the inverse, Eqs.~(\ref{invers1}) and (\ref{invers2}), the
tensor field $\omega^{ij}$, being the (unique) inverse of
$\omega_{ij}$, is given by
\begin{eqnarray}
\label{A} \omega^{ik}\;\omega_{kj}
(-1)^{\epsilon_k+\epsilon(\omega)(\epsilon_i+\epsilon_k)}=
\delta^i_j,\quad
(-1)^{\epsilon_i+\epsilon(\omega)(\epsilon_i+\epsilon_k)}
\omega_{ik}\;\omega^{kj}=\delta^j_i.
\end{eqnarray}
The inverse tensor field $\omega^{ij}$ has the following symmetry
property:
\begin{eqnarray}
\label{A11}
\omega^{ij}=-(-1)^{\epsilon_i\epsilon_j+\epsilon(\omega)}\omega^{ji},\quad
\epsilon(\omega^{ij})= \epsilon(\omega)+\epsilon_i+\epsilon_j.
\end{eqnarray}
In terms of the inverse tensor field $\omega^{ij}$, the relations
(\ref{A12}) can be rewritten as follows:
\begin{eqnarray}
\label{A7} \omega^{in}\partial_n\omega^{jk}
(-1)^{(\epsilon_i+\epsilon(\omega))(\epsilon_k+\epsilon(\omega))}
+cycle (i,j,k)\equiv 0\,.
\end{eqnarray}

With the help of $\omega^{ij}=\omega^{ij}(x)$ one can introduce
the even (resp. odd) Poisson bracket,
\begin{eqnarray}
\label{A1} (A,B)=\frac{\partial_r A}{\partial x^i}\;
(-1)^{\epsilon(\omega)\epsilon_i}\omega^{ij}\; \frac{\partial
B}{\partial x^j},\quad
\epsilon((A,B))=\epsilon(A)+\epsilon(B)+\epsilon(\omega)\,,
\end{eqnarray}
which is invariant under transformations $(x)\rightarrow ({\bar
x})$ of the coordinates, $({\bar A},{\bar B})=(A,B)$. One easily
verifies that all the properties which are required for a bilinear
form to be a Poisson bracket are fulfilled. In particular, the
relation (\ref{A7}) is just the (super) Jacobi identity for the
Poisson bracket.

Supermanifolds equipped with a tensor field obeying the properties
(\ref{A11}) and (\ref{A7}) are called even (odd) Poisson
supermanifolds. From the above considerations it follows that, as
in the case of ordinary differential geometry, there exists an
one-to-one correspondence between even (odd) non-degenerate
Poisson supermanifolds and even (odd) symplectic supermanifolds.

\pagebreak

\begin{center}
{\bf 5. FEDOSOV SUPERMANIFOLDS}
\end{center}

Suppose now we are given an even (odd) symplectic supermanifold,
$(M,\omega)$. Let $\nabla$ (or $\Gamma$) be a covariant derivative
(connection) on $M$ which preserves the 2-form $\omega$,
$\omega\nabla=0$.  In a coordinate basis this requirement reads
\begin{eqnarray}
\label{covomiv} \omega_{ij,k}-\omega_{im}\Gamma^m_{\;\;jk}+
\omega_{jm}\Gamma^m_{\;\;ik}(-1)^{\epsilon_i\epsilon_j}=0.
\end{eqnarray}
If, in addition, $\Gamma$ is symmetric then we have an even (odd)
symplectic connection (or symplectic covariant derivative) on $M$.
Now, an even (odd) Fedosov supermanifold $(M,\omega,\Gamma)$ is
defined as an even (odd) symplectic supermanifold with a given
even (odd) symplectic connection.

Let us introduce the curvature tensor of an even (odd) symplectic
connection,
\begin{eqnarray}
\label{Rs}
R_{ijkl}=\omega_{in}R^n_{\;\;jkl},\quad
\epsilon(R_{ijkl})=\epsilon(\omega)+\epsilon_i+
\epsilon_j+\epsilon_k+\epsilon_l,
\end{eqnarray}
where $R^n_{\;\;jkl}$ is given by (\ref{R}). This leads to the
following representation,
\begin{eqnarray}
\label{Rse} R_{imjk}=-\omega_{in}\Gamma^n_{\;\;mj,k}+
\omega_{in}\Gamma^n_{\;\;mk,j}(-1)^{\epsilon_j\epsilon_k}+
\Gamma_{ijn}\Gamma^n_{\;\;mk}(-1)^{\epsilon_j\epsilon_m}-
\Gamma_{ikn}\Gamma^n_{\;\;mj}
(-1)^{\epsilon_k(\epsilon_m+\epsilon_j)}\,,
\end{eqnarray}
where we used the notation
\begin{eqnarray}
\label{G} \Gamma_{ijk}=\omega_{in}\Gamma^n_{\;\;jk},\quad
\epsilon(\Gamma_{ijk})=\epsilon(\omega)+
\epsilon_i+\epsilon_j+\epsilon_k\,.
\end{eqnarray}
Using this, the relation (\ref{covomiv}) reads
\begin{eqnarray}
\label{covom}
\omega_{ij,k}=\Gamma_{ijk}-
\Gamma_{jik}(-1)^{\epsilon_i\epsilon_j}.
\end{eqnarray}
Furthermore, from Eq.~(\ref{R}) it is obvious that
\begin{eqnarray}
\label{Rans} R_{ijkl}=-(-1)^{\epsilon_k\epsilon_l}R_{ijlk},
\end{eqnarray}
and, using (\ref{Rs}) and (\ref{Rjac}), one deduces the (super)
Jacobi identity for $R_{ijkl}$,
\begin{eqnarray}
\label{Rjac1} (-1)^{\epsilon_j\epsilon_l}R_{ijkl}
+(-1)^{\epsilon_l\epsilon_k}R_{iljk}
+(-1)^{\epsilon_k\epsilon_j}R_{iklj}=0\,.
\end{eqnarray}

In addition, the curvature tensor $R_{ijkl}$ is (generalized)
symmetric w.r.t. the first two indices,
\begin{eqnarray}
\label{Ras}
R_{ijkl}=(-1)^{\epsilon_i\epsilon_j}R_{jilk}.
\end{eqnarray}
In order to prove this, let us consider
\begin{eqnarray}
\label{com}
\omega_{ij,kl}=\Gamma_{ijk,l}-
\Gamma_{jik,l}(-1)^{\epsilon_i\epsilon_j}.
\end{eqnarray}
Then, using the relations
\begin{eqnarray}
\label{G1} \Gamma_{ijk,l}=\omega_{in}\Gamma^n_{\;\;jk,l}
+\omega_{in,l}\Gamma^n_{\;\;jk}
(-1)^{(\epsilon_n+\epsilon_j+\epsilon_k)\epsilon_l}
\end{eqnarray}
and the definitions (\ref{Rse}) and (\ref{covom}), we get
\begin{eqnarray}
\label{com1}
\nonumber
0&=&\omega_{ij,kl}-(-1)^{\epsilon_k\epsilon_l}\omega_{ij,lk}\\
\nonumber
&=&\Gamma_{ijk,l}-
\Gamma_{jik,l}(-1)^{\epsilon_i\epsilon_j}
-\Gamma_{ijl,k}(-1)^{\epsilon_k\epsilon_l}+
\Gamma_{jil,k}(-1)^{\epsilon_i\epsilon_j+\epsilon_k\epsilon_l}\\
&=&-R_{ijkl}+(-1)^{\epsilon_i\epsilon_j}R_{jikl}.
\end{eqnarray}

For any even (odd) symplectic connection there holds the identity
\begin{eqnarray}
\label{Rjac2}
(-1)^{\epsilon_i\epsilon_l}R_{ijkl}
+(-1)^{\epsilon_l\epsilon_k+\epsilon_l\epsilon_j}R_{lijk}
+(-1)^{\epsilon_k\epsilon_j+\epsilon_l\epsilon_j+\epsilon_i\epsilon_k}
R_{klij}+
(-1)^{\epsilon_i\epsilon_j+\epsilon_i\epsilon_k}R_{jkli}=0.
\end{eqnarray}
This is proved by using the Jacobi identity (\ref{Rjac1}) together
with a cyclic change of the indices:
\begin{eqnarray}
\label{Ap1}
(-1)^{\epsilon_j\epsilon_l}R_{ijkl}
+(-1)^{\epsilon_l\epsilon_k}R_{iljk}
+(-1)^{\epsilon_k\epsilon_j}R_{iklj}=0,\\
\label{Ap2}
(-1)^{\epsilon_i\epsilon_k}R_{lijk}
+(-1)^{\epsilon_j\epsilon_k}R_{lkij}
+(-1)^{\epsilon_i\epsilon_j}R_{ljki}=0,\\
\label{Ap3}
(-1)^{\epsilon_j\epsilon_l}R_{klij}
+(-1)^{\epsilon_i\epsilon_j}R_{kjli}
+(-1)^{\epsilon_i\epsilon_l}R_{kijl}=0,\\
\label{Ap4}
(-1)^{\epsilon_i\epsilon_k}R_{jkli}
+(-1)^{\epsilon_i\epsilon_l}R_{jikl}
+(-1)^{\epsilon_l\epsilon_k}R_{jlik}=0.
\end{eqnarray}
Now, multiplying Eq. (\ref{Ap1}) by the factor
$(-1)^{\epsilon_i\epsilon_l+\epsilon_j\epsilon_l}$ and Eq.
(\ref{Ap3}) by the factor
$(-1)^{\epsilon_i\epsilon_k+\epsilon_j\epsilon_k}$ and summing the
obtained results, one gets the identity (\ref{Rjac2}). The same is
obtained by multiplying Eq. (\ref{Ap2}) by the factor
$(-1)^{\epsilon_i\epsilon_k+\epsilon_k\epsilon_l}$ and Eq.
(\ref{Ap4}) by the factor
$(-1)^{\epsilon_i\epsilon_j+\epsilon_j\epsilon_l}$ and then
summing the results. Moreover, any other combination of Eqs.
(\ref{Ap1})-(\ref{Ap4}) containing four components of the
symplectic curvature with a cyclic permutation of all the indices
are reduced to the identities (\ref{Rjac2}).
In the case of ordinary Fedosov manifolds, i.e., when all the
variables $x^i$ are even ($\epsilon_i=0$), Eq. (\ref{Rjac2})
obtains the symmetric form \cite{fm},
\begin{eqnarray}
\label{Rjac3}
R_{ijkl} + R_{lijk} +R_{klij} +R_{jkli}=0.
\end{eqnarray}

In the identity (\ref{Rjac2}) the components of the symplectic
curvature tensor occur with cyclic permutations of all the
indices. However, the pre-factors depending on the Grassmann
parities of indices are not obtained by cyclic permutation. One
may consider this as an unexpected result, but the Jacobi identity
(\ref{Rjac1}) obeys this property only w.r.t. the last three
indices and, in addition, also the (anti) symmetry properties
(\ref{Rans}) and (\ref{Ras}) do not fulfill this requirement.

\begin{center}
{\bf 6. RICCI TENSOR}
\end{center}

Having  the curvature tensor, $R_{ijkl}$, and the tensor field
$\omega^{ij}$, with allowance made for the symmetry properties of
these tensors, (\ref{A11}), (\ref{Rans}) and (\ref{Ras}),
one can define the following three different tensor fields of type
$(0,2)$,
\begin{eqnarray}
\label{R1}
&&R_{ij}=\omega^{kn}R_{nkij}
 (-1)^{(\epsilon(\omega)+1)(\epsilon_k+\epsilon_n)}
 \qquad\;=\;R^k_{\;\;kij}\;(-1)^{\epsilon_k}
 \,,\\
\label{R2} &&K_{ij}=
\omega^{kn}R_{nikj}
 (-1)^{\epsilon_i\epsilon_k+(\epsilon(\omega)+1)(\epsilon_k+\epsilon_n)}
 \;=\;R^k_{\;\;ikj}\;(-1)^{\epsilon_k(\epsilon_i+1)}\,,\\
\label{R3}
&&Q_{ij}=\omega^{kn}R_{ijnk}
 (-1)^{(\epsilon_i+\epsilon_j)(\epsilon_k+\epsilon_n)+
 (\epsilon(\omega)+1)(\epsilon_k+\epsilon_n)}
 \,,\\
\nonumber &&\epsilon(R_{ij})=\epsilon(K_{ij})=\epsilon(Q_{ij})=
\epsilon_i + \epsilon_j\,,
\end{eqnarray}
where, obviously, $R_{ij}$ and $K_{ij}$ do not depend on $\omega$.

From the definitions (\ref{R1}), (\ref{R3}) and the symmetry
properties of $R_{ijkl}$, it follows immediately  that for any
symplectic connection one has
\begin{eqnarray}
\label{R1s}
&&R_{ij}=-(-1)^{\epsilon_i\epsilon_j}R_{ji},\\
\label{R3s}
&&Q_{ij}=(-1)^{\epsilon_i\epsilon_j}Q_{ji}.
\end{eqnarray}
Therefore, on any even Fedosov supermanifold the tensor $R_{ij}$
($Q_{ij}$) equals to zero (is non-trivial),  while on any odd
Fedosov supermanifold this tensor is non-trivial (equals to zero).
Indeed, using the symmetry properties of the tensor fields
$\omega^{ij}$ and $R_{ijkl}$, one obtains the relations
\begin{eqnarray}
\label{R1p}
[1+(-1)^{\epsilon(\omega)}]R_{ij}=0,\\
\label{R3p}
[1-(-1)^{\epsilon(\omega)}]Q_{ij}=0.
\end{eqnarray}

In addition, for the tensor fields (\ref{R1}) -- (\ref{R3}) there
exists a relation which follows from the identity (\ref{Rjac2}).
Indeed, multiplying the equation (\ref{Rjac2}) by the factor $
(-1)^{\epsilon_i\epsilon_l+(\epsilon(\omega)+1)(\epsilon_i+\epsilon_j)}
$ and by the tensor $\omega^{ji}$, with allowance made for
summation over indices $i,j$ and for the definitions (\ref{R1}) --
(\ref{R3}), we obtain
\begin{eqnarray}
\label{Rl1}
R_{ij}+Q_{ij}+(-1)^{\epsilon_i\epsilon_j}K_{ji}+
(-1)^{\epsilon(\omega)}K_{ij}=0.
\end{eqnarray}
A second independent relation can be derived from (\ref{Rjac2}) by
multiplying with the tensor $\omega^{ki}$ and the factor $
(-1)^{\epsilon_i\epsilon_l+\epsilon_j\epsilon_k
(\epsilon(\omega)+1)(\epsilon_i+\epsilon_j)} $; after subsequent
summation over the indices $i,k$ this leads to the following
result:
\begin{eqnarray}
\label{Rl2}
[1+(-1)^{\epsilon(\omega)}]\left(K_{ij}-(-1)^{\epsilon_i\epsilon_j}K_{ji}\right)=0.
\end{eqnarray}

From the relations (\ref{R1p}), (\ref{R3p}) and (\ref{Rl2}) one
concludes: For any even symplectic connection we obtain
\begin{eqnarray}
\label{Rl3} K_{ij}=(-1)^{\epsilon_i\epsilon_j}K_{ji}, \quad
R_{ij}=0,\quad Q_{ij}=-2K_{ij},
\end{eqnarray}
while for any odd symplectic connection we have
\begin{eqnarray}
\label{Rl4} Q_{ij}=0,\quad
R_{ij}=K_{ij}-(-1)^{\epsilon_i\epsilon_j}K_{ji}.
\end{eqnarray}
Therefore, the tensor field $K_{ij}$ should be considered as the
only independent second-rank tensor which can be constructed from
the symplectic curvature. We refer to $K_{ij}$ as the Ricci tensor
of an even (odd) Fedosov supermanifold.

Let us define the scalar curvature $K$ by the formula
\begin{eqnarray}
\label{Rsc} K=\omega^{ji}K_{ij}(-1)^{\epsilon_i+\epsilon_j}=
\omega^{ji}\omega^{kn}R_{nikj}
(-1)^{\epsilon_i+\epsilon_j+\epsilon_i\epsilon_k+
(\epsilon_k+\epsilon_n)(\epsilon(\omega)+1)}.
\end{eqnarray}
From the symmetry properties of $R_{ijkl}$ and $\omega^{ij}$,
it follows
that on any Fedosov supermanifold one has
\begin{eqnarray}
\label{Rsc1} [1+(-1)^{\epsilon(\omega)}]K=0.
\end{eqnarray}
Therefore, as is the case for ordinary Fedosov manifolds
\cite{fm}, for any even symplectic connection the scalar curvature
necessarily vanishes. But the situation becomes different for odd
Fedosov supermanifolds where no restriction on the scalar
curvature occurs. Therefore, in contrast to both the usual Fedosov
manifolds and the even Fedosov supermanifolds, any odd Fedosov
supermanifolds can be characterized by the scalar curvature as
additional geometrical structure.


\begin{center}
{\bf 7. SUMMARY}
\end{center}

We have considered  some properties of tensor fields defined on
supermanifolds $M$. It was shown that only the generalized
(anti)symmetry of tensor fields has an invariant meaning, and that
differential geometry on supermanifolds should be constructed in
terms of such tensor fields.

Any supermanifold $M$ can be equipped with a symmetric connection
$\Gamma$ (covariant derivative $\nabla$). The Riemannian tensor
$R^i_{\;\;jkl}$ corresponding to this symmetric connection
$\Gamma$ satisfies both the Jacobi identity and the Bianchi
identity.

Any supermanifold $M$ of an even dimension can be endowed with an
even (odd) 2-form $\omega$. If this 2-form is non-degenerate and
closed, the pair $(M,\omega)$ defines an even (odd) symplectic
supermanifold. In a coordinate basis on the supermanifold $M$, the
2-form $\omega$ is described by a second-rank tensor field
$\omega_{ij}$ obeying the property of generalized antisymmetry in
both the even and odd cases. In its turn, the tensor field
$\omega^{ij}$, being inverse to $\omega_{ij}$, obeys the property
of generalized antisymmetry in the even case, while in the odd
case it has the property of generalized symmetry. The tensor
$\omega^{ij}$ defines the even (odd) Poisson bracket on a
supermanifold $M$. The Jacobi identity for the even (odd) Poisson
bracket follows from the closure of the 2-form $\omega$.
Supermanifolds equipped with an even (odd) non-degenerate Poisson
structure $\omega^{ij}$ are called even (odd) Poisson
supermanifolds. Therefore, there exists an one-to-one
correspondence between an even (odd) symplectic supermanifold and
the corresponding even (odd) Poisson supermanifold.

Any even (odd) symplectic supermanifold can be equipped with a
symmetric connection respecting the given symplectic structure.
Such a symmetric connection is called a symplectic connection. The
triplet $(M,\omega,\Gamma)$ is called an even (odd) Fedosov
supermanifold. The curvature tensor $R_{ijkl}$ of a symplectic
connection obeys the property of generalized symmetry with respect
to the first two indices, and the property of generalized
antisymmetry with respect to the last two indices. The tensor
$R_{ijkl}$ satisfies the Jacobi identity and the specific (for the
symplectic geometry) identity (see (\ref{Rjac2})) containing the
sum of components of this tensor with a cyclic permutation of all
the indices, which, however, does not  (!) contain cyclic permuted
factors depending on the Grassmann parities of the indices.

On any even (odd) Fedosov manifold, the Ricci tensor $K_{ij}$ can
be defined. In the even case, the Ricci tensor obeys the property
of generalized symmetry and gives a trivial result for the scalar
curvature. On the contrary, in the odd case the scalar curvature,
in general, is nontrivial.

{\sc Acknowledgements:}~The authors are grateful to S. Bellucci, A.V.
Galajinsky, M. Henneaux, A.P. Nersessian and D.V. Vassilevich for
stimulating discussions. The work was supported by
Deutsche Forschungsgemeinschaft (DFG) grant GE 696/7-1.
The  work of P.M.L. was also supported under
the projects DFG 436 RUS
113/669,  Russian Foundation for Basic Research (RFBR)
02-02-04002, 03-02-16193 and the President grant 1252.2003.2 for
supporting leading scientific schools.

\end{document}